# Retrieving Instantaneous Field of View and Geophysical Information for Atmospheric Limb Sounding with USGNC Near Real-Time Orbit Data

Laura Cui, Montgomery Blair High School

SEAP Mentor: Dr. Andrew Stephan, U.S. Naval Research Laboratory

August 18, 2017

# 1 INTRODUCTION

The Earth's ionosphere is an energetic region of the upper atmosphere, characterized by a partially ionized plasma which forms when neutral particles are ionized by high-energy solar or cosmic radiation. The ionosphere hosts a variety of dynamic processes which drive strong gradients in density and electric field, which manifest as disturbances such as the equatorial arcs. These gradients may also form smaller secondary perturbations and plasma irregularities, which interact with electromagnet waves passing through the region and can distort or otherwise disrupt electromagnet waves including radio and satellite communications[1].

The Limb-imaging Ionospheric and Thermospheric Extreme-ultraviolet Spectrograph (LITES) experiment is one of thirteen instruments aboard the Space Test Program Houston 5 (STP-H5) payload on the International Space Station (ISS). The mission launched February of 2017 and will operate for at least two years. The primary science goal of the LITES experiment is to investigate ionospheric structures and variability relevant to the global ionosphere, along with the complementary GPS Radio Occultation and Ultraviolet Photometry – Colocated (GROUP-C) experiment. Since the ISS has an orbital inclination of 51.6° which combined with its altitude of about 410 km enables middle- and low-latitude measurements from slightly above the peak region of the ionosphere[2][3].

# 2 REMOTE SENSING

## 2.1 Passive Remote Sensing

Atmospheric sounding provides an opportunity to measure and characterize vertical distribution of density and other properties in the atmospheric column. Although active sensors such as radar instruments are able to reliably measure physical properties of the atmosphere, they do not directly determine density or composition. It has been previously demonstrated that density and species concentration can be measured using naturally occurring emissions in the ionosphere, such as the O 135.6 nm and $O^+$ 83.4 and 61.6 nm emissions[4][5]. LITES is a passive remote sensor to measure these and other emissions in the extreme and far ultraviolet (EUV/FUV) spectrum[3].

## 2.2 Limb Viewing Geometry

A limb-viewing instrument looks towards the edge of the atmosphere visible above Earth's horizon, producing a vertical field of view. In contrast, a nadir-viewing instrument looks down towards the Earth's surface. A nadir view produces high horizontal spatial resolution, but offers no vertical information. LITES' field of view therefore allows it to more reliably measure vertical density profiles in the upper atmosphere[2].

# 3 FIELD OF VIEW

## 3.1 Instrument Configuration

LITES has a 10° by 10° field of view which is collapsed horizontally, combining all information from a given altitude. The instrument is installed such it looks in the wake of the ISS and about 14.5° downwards, enabling it to image altitudes ranging from about 350 km to 150 km, assuming it experiences an average pitch of about -2.6° aboard the ISS. The actual geophysical information captured by LITES is highly dependent on the pitch of the ISS, which directly affects the viewing altitude and geometry of the instrument[2][3].

## 3.2 ISS Data Format

The ISS Guidance, Navigation, and Control (GNC) System consists of multiple control systems, including the U.S. GNC System and the Russian Orbital Segment Motion Control System (ROS MCS). U.S. GNC is the primary system for the attitude determination function of the ISS GNC System and relies primarily on Global Positioning System (GPS) interferometry to update position, velocity, and attitude every 10 seconds. U.S. GNC also includes a set of rate gyro assemblies which measure variations in laser beams to sense changes in attitude with higher cadence. U.S. GNC data is consistently available in real time[6].

U.S. GNC data is processed and represented in several distinct coordinate systems, including two of interest. The first is the True of Date (TR), Cartesian Coordinate System, an example of an Earth-centered inertial (ECI) system. TR is a quasi-inertial right-handed Cartesian coordinate system setting the origin at the center of the Earth, with the X axis as the true (current) vernal equinox and the Z axis as the true rotational axis. TR is useful for computing the location of stars and the Sun relative to the field of view[7].

The Local Orbital: Local Vertical Local Horizontal (LVLH) system is a rotating right-handed Cartesian coordinate system setting the origin at the center of the ISS, with the Y axis opposite to the orbit momentum vector and normal to the orbit plane, and the Z axis as the geocentric radius vector to the ISS and positive towards Earth's center. The X axis is approximately the velocity vector during normal ISS operation[7]. LVLH simplifies the viewing geometry of the LITES instrument and allows the look vector to be transformed to an ECI coordinate system to determine the field of view.

## 3.3 Attitude Quaternions

Quaternions provide an elegant and less resource-expensive method of spacecraft attitude determination compared to the traditional Euler angles representation. Three parameters are required to define a rotational orientation in three-dimensional space, one for each axis. It is common to express an orientation in terms of a rotation matrix, which can be directly applied as a rotation operation. An arbitrary orientation can be expressed using the following elementary rotation matrices:

$$R^x(\theta) = \begin{bmatrix} 1 & 0 & 0 \\ 0 & \cos\theta & \sin\theta \\ 0 & -\sin\theta & \cos\theta \end{bmatrix} \quad R^y(\theta) = \begin{bmatrix} \cos\theta & 0 & -\sin\theta \\ 0 & 1 & 0 \\ \sin\theta & 0 & \cos\theta \end{bmatrix} \quad R^z(\theta) = \begin{bmatrix} \cos\theta & \sin\theta & 0 \\ -\sin\theta & \cos\theta & 0 \\ 0 & 0 & 1 \end{bmatrix} \tag{3.1}$$

These matrices can be combined into a single rotation matrix with a new "Euler axis" or axis of rotation. However, using this method to determine a representation for an arbitrary rotation is computationally intense. Alternatively, rotational orientation can be described using a quaternion, which has a scalar component *s* and vector component ***v***, represented as four numbers:

$$q = \begin{bmatrix} s \\ \vec{v} \end{bmatrix} = \begin{bmatrix} s \\ v_x \\ v_y \\ v_z \end{bmatrix} = \begin{bmatrix} q_s \\ q_x \\ q_y \\ q_z \end{bmatrix} \tag{3.2}$$

where

$$\begin{bmatrix} q_s \\ q_x \\ q_y \\ q_z \end{bmatrix} = \begin{bmatrix} \cos\frac{\theta}{2} \\ ||\vec{e}|| \cdot \sin\frac{\theta}{2} \end{bmatrix} \tag{3.3}$$

and ***e*** represents the Euler axis. We can then recover the rotation matrix:

$$\mathrm{R} = \begin{bmatrix} 1-2s(q_y^2+q_z^2) & 2s(q_xq_y-q_zq_s) & 2s(q_xq_z+q_yq_s) \\ 2s(q_xq_y-q_zq_s) & 1-2s(q_x^2+q_z^2) & 2s(q_yq_z+q_xq_s) \\ 2s(q_xq_z-q_yq_s) & 2s(q_yq_z+q_xq_s) & 1-2s(q_x^2+q_y^2) \end{bmatrix}$$

$$s = ||q||^{-2} \tag{3.4}$$

This allows us to perform transformations between different reference coordinate systems[8].

### 3.4 Estimating Field of View

Given an angle $\theta$ representing the orientation of the LITES look vector relative to the $-X_{LVLH}$ axis, we can represent the look vector as:

$$\vec{l} = \begin{bmatrix} \cos(\theta), \\ 0, \\ \sin(\theta) \end{bmatrix} \quad (3.5)$$

We can then use the LVLH attitude quaternion, as well as the position and velocity vector, to recover the transformation between the ECI and LVLH coordinate systems. (We can also use the LVLH attitude quaternion to determine the roll, pitch, and yaw of the ISS using triangle geometry.) Matrix multiplication yields the desired vector in ECI coordinates.

### 3.5 Incorporating Stellar Observation

Recall that LITES has a 10° by 10° field of view which is collapsed horizontally to produce a vertical profile. Since stars have a continuous spectrum, they appear as a horizontal line in the resolved spectrogram. Stars generally pass through the field of view at a slight angle due to the orientation of the instrument, which we can use to determine the orientation of the field of view of the sensor.

By fitting the star's observed peak to a Gaussian distribution, we can map the observed motion of the star through the field of view. To determine the rate of motion, we can perform linear regression on the vertical position of the star with respect to time. We expect that $r \approx 1$ due to the consistent motion of the ISS.

We can also generate a table of expected vertical motion using our original estimate of the LITES field of view and coordinates from a standard star catalogue. We can then compare the expected and observed rate of motion, as well as the vertical position of the star in the field of view, to develop a transformation mapping our estimate to the actual field of view.

It may be important to recognize that we are working with the TR coordinate system rather than a truly inertial coordinate system such as J2000, and to consider possible implications of the difference. The TR coordinate system uses "true of date" references, which vary slightly from those of the J2000 epoch due to the effects of precession. Therefore, there may be differences between the calculated and observed look vector for stellar coordinates given relative to the J2000 epoch. However, the rate of axial precession can be determined using the IAU 2000A model as well as long-range interferometry techniques. The model yields a slowly diverging rate of accumulated precession of approximately 5,028.796195" per Julian century or 0.01396887831" per Julian year, centered at the J2000 epoch. Earth's axial precession can therefore be resolved for the J2000 epoch[9][10].

## 4 CONCLUSION

Determining the field of view of the LITES instrument allows geophysical and altitude information to be retrieved, a crucial step in reconstructing vertical structure and density profiles in the upper atmosphere. Since the LITES viewing geometry is especially sensitive to change in the pitch of the ISS, it is necessary to process the actual orbit data rather than relying on ephemeris tables, which do not accurately reflect the quasi-sinusoidal nature of variations in the attitude of the ISS.

Incorporating stellar observations into the vector geometry of the instrument configuration allow the field of view to be determined with sufficient accuracy for the purposes of the LITES experiment. This estimate serves as a necessary first step for analysis, along with other data pre-processing procedures such as for noise or optical distortion.

# 5 ACKNOWLEDGEMENTS

This work was made possible by the guidance of my mentor, Dr. Andrew Stephan, as well as the work of Dr. Scott Budzien and the LITES team at the U.S. Naval Research Laboratory and the University of Massachusetts Lowell, with support from the Chief of Naval Research. In addition, the opportunity to perform this research was provided by the American Society for Engineering Education and the Science and Engineering Apprenticeship Program.